\documentclass[twocolumn,times]{aastex7}
\usepackage{caption}
\usepackage{color}
\usepackage{soul}

\usepackage{makecell}

\begin{document}

\title{Molecular Gas Structure and Star Formation Diversity in Stephan's Quintet Revealed by ACA CO(1-0) Mapping}

\author[orcid=0009-0006-4760-9771]{Misaki Yamamoto}
\affiliation{Department of Physics, Graduate School of Science, Osaka Metropolitan University, 3-3-138 Sugimoto, Sumiyoshi-ku, Osaka, 558-8585, Japan}
\email{sq25473q@st.omu.ac.jp} 

\author[orcid=0000-0002-8868-1255]{Fumiya Maeda}
\affiliation{Research Center for Physics and Mathematics, Osaka Electro-Communication University, 18-8 Hatsucho, Neyagawa, Osaka 572-8530, Japan}
\email{fmaeda@osakac.ac.jp}  

\author[orcid=0000-0002-3373-6538]{Kazuyuki Muraoka}
\affiliation{Department of Physics, Graduate School of Science, Osaka Metropolitan University, 3-3-138 Sugimoto, Sumiyoshi-ku, Osaka, 558-8585, Japan}
\email{kmuraoka@omu.ac.jp}  

\author[orcid=0000-0002-1639-1515]{Fumi Egusa}
\affiliation{Institute of Astronomy, Graduate School of Science, The University of Tokyo, 2-21-1 Osawa, Mitaka, Tokyo 181-0015, Japan}
\email{fegusa@ioa.s.u-tokyo.ac.jp}

\author[orcid=0009-0007-2493-0973]{Shinya Komugi}
\affiliation{Division of Liberal Arts, Kogakuin University, 2665-1 Nakano-cho, Hachioji, Tokyo 192-0015, Japan}
\email{}

\author[0000-0001-6469-8725]{Bunyo Hatsukade}
\affiliation{National Astronomical Observatory of Japan, 2-21-1 Osawa, Mitaka, Tokyo 181-8588, Japan}
\affiliation{Graduate University for Advanced Studies, SOKENDAI, Osawa, Mitaka, Tokyo 181-8588, Japan}
\affiliation{Department of Astronomy, Graduate School of Science, The University of Tokyo, 7-3-1 Hongo, Bunkyo-ku, Tokyo 133-0033, Japan}
\email{}

\author[0000-0002-2699-4862]{Hiroyuki Kaneko}
\affiliation{College of Creative Studies, Niigata University, 8050
Ikarashi 2-no-cho,
Nishi-ku, Niigata, Niigata 950-2181, Japan}
\email{kaneko.hiroyuki.astro@gmail.com}

\author[orcid=0000-0003-3990-1204]{Masato I.N. Kobayashi} 
\affiliation{Meta-hierarchy Dynamics Unit, National Institute for Fusion Science, 322-6 Oroshi-Cho, Toki City, Gifu 509-5292, Japan} \email{kobayashi.masato@nifs.ac.jp}

\author[orcid=0000-0002-4052-2394]{Kotaro Kohno}
\affiliation{Institute of Astronomy, Graduate School of Science, The University of Tokyo, 2-21-1 Osawa, Mitaka, Tokyo 181-0015, Japan}
\affiliation{Research Center for the Early Universe, Graduate School of Science, The University of Tokyo, 7-3-1 Hongo, Bunkyo-ku, Tokyo 113-0033, Japan}
\email{kkohno@ioa.s.u-tokyo.ac.jp}

\author[orcid=0000-0002-4098-8100]{Ayu Konishi}
\affiliation{Department of Physics, Graduate School of Science, Osaka Metropolitan University, 3-3-138 Sugimoto, Sumiyoshi-ku, Osaka, 558-8585, Japan}
\affiliation{Astronomical Institute, Tohoku University, 6-3, Aramaki, Aoba-ku, Sendai, Miyagi 980-8578, Japan}
\email{ayu.konishi@astr.tohoku.ac.jp <ayu.konishi@astr.tohoku.ac.jp}

\author[orcid=0000-0003-3932-0952, gname=Kana,sname='Morokuma-Matsui']{Kana Morokuma-Matsui}
\affiliation{Institute of Astronomy, Graduate School of Science, The University of Tokyo, 2-21-1 Osawa, Mitaka, Tokyo 181-0015, Japan}
\email{kanamoro@ioa.s.u-tokyo.ac.jp}  

\author[orcid=0000-0002-6939-0372]{Kouichiro Nakanishi}
\affiliation{National Astronomical Observatory of Japan, 2-21-1 Osawa,
Mitaka, Tokyo 181-8588, Japan}
\affiliation{Graduate Institute for Advanced Studies, SOKENDAI, Osawa, Mitaka, Tokyo 181-8588, Japan}
\email{}

\author[orcid=0000-0003-3844-1517]{Kouji Ohta}
\affiliation{Department of Astronomy, Kyoto University,  Kyoto,  606-8502, Japan}
\email{ohta@kusastro.kyoto-u.ac.jp}

\begin{abstract}
We present $^{12}$CO(1--0) mapping across the entire system of Stephan’s Quintet, a well-known compact galaxy group, observed by Atacama Compact Array (7\,m array + Total Power) of the Atacama Large Millimeter/submillimeter Array. These observations provide the first large-scale ($137\,\mathrm{kpc}\times119\,\mathrm{kpc}$), spatially resolved ($\sim$5.5\,$\mathrm{kpc}$) molecular gas map of a compact group.
Our CO map revealed that most of the molecular gas resides in the disk of the member galaxy NGC~7319 and in the intergalactic regions, including components along the shocked filament and the optically identified tidal tail extending from NGC~7319. 
Along the tidal tail and its surroundings, we found not only an extended molecular gas component but also four discrete CO clumps, with velocity dispersions of $\sim$10–30~$\mathrm{km\,s^{-1}}$ and molecular gas masses of order $10^7$–$10^8\,M_\odot$. 
Three of these clumps spatially overlap with H\,{\sc i}, whereas the remaining clump shows no associated H\,{\sc i} or counterparts at optical and infrared wavelengths. 
Using star formation rates derived from H$\alpha$ luminosities of H\,{\sc ii} regions, we found that star formation efficiencies (SFEs) span $\sim$2.2\,dex ($\sim$0.02--4\,Gyr$^{-1}$) and negatively correlate with CO velocity dispersion.
While regions with small velocity dispersion exhibit SFEs comparable to those of nearby disk galaxies, those with large velocity dispersion ($\sim$50-150$\,\mathrm{km\,s^{-1}}$) around the shocked filament show strongly suppressed star formation. These results suggest that turbulence plays a significant role in regulating star formation in interacting systems.
\end{abstract}



\section{Introduction} 
\label{sec: intro}
Galaxy interactions play a crucial role in galaxy evolution.
For example, they are known to drive morphological changes, including the excitation of spiral arms and bar structures \citep[e.g.,][]{Toomre_1972ApJ,Noguchi_1987MNRAS,Gerin_1990A&A,Dobbs_2010MNRAS,Lokas_2014MNRAS,Pettitt_2018MNRAS}, and the evolution of disk galaxies into elliptical systems through mergers \citep[e.g.,][]{Toomre_1972ApJ,Burkert_2003LNP,Bournaud_elliptical_2007A&A}.
Galaxy interactions can change the physical state of gas within galaxies and/or remove gas from them, giving rise to contrasting star-formation outcomes, including starbursts \citep[e.g.,][]{Larson_1978ApJ,Barton_2000ApJ,Matteo_2008A&A,Patton_2011MNRAS} as well as, conversely, the suppression of star formation \citep[e.g.,][]{Robotham_2013MNRAS,Davies_2015MNRAS,Alatalo_2015ApJ}.
Clarifying how interactions modify the physical state of gas and regulate star formation is therefore central to understanding the diverse evolutionary pathways of interacting galaxies.

Compact galaxy groups are thought to experience frequent interactions and mergers due to their close proximity.
Such systems provide ideal laboratories for studying these interaction-driven processes.
They often contain large amounts of intergalactic gas, reflecting their interaction histories, which in turn offer valuable clues to the evolution of the member galaxies.
Furthermore, young star-forming regions have been identified within intergalactic gas \citep[e.g.,][]{braine_abundant_2001,Lisenfeld_2002AA,Duc_2007A&A,Boquien_2011A&A}, providing evidence that galaxy interactions can trigger star formation outside galactic disks.
Therefore, studying the entire group, including the intergalactic medium, is essential not only for investigating the evolutionary histories of member galaxies, but also for assessing how galaxy interactions influence the evolution of gas on group-wide scales.
However, most previous studies of compact groups have focused on H\,{\sc i} \citep[e.g.,][]{Verdes-Montenegro_2001A&A,Williams_1991AJ,Williams_2002AJ,Serra_2013MNRAS,Jones_2023A&A}, while spatially resolved observations of molecular gas remain scarce, leaving its global distribution, physical properties, and link to star formation poorly constrained.

\begin{figure}[t!]
 \begin{center}
  \includegraphics[width=85mm]{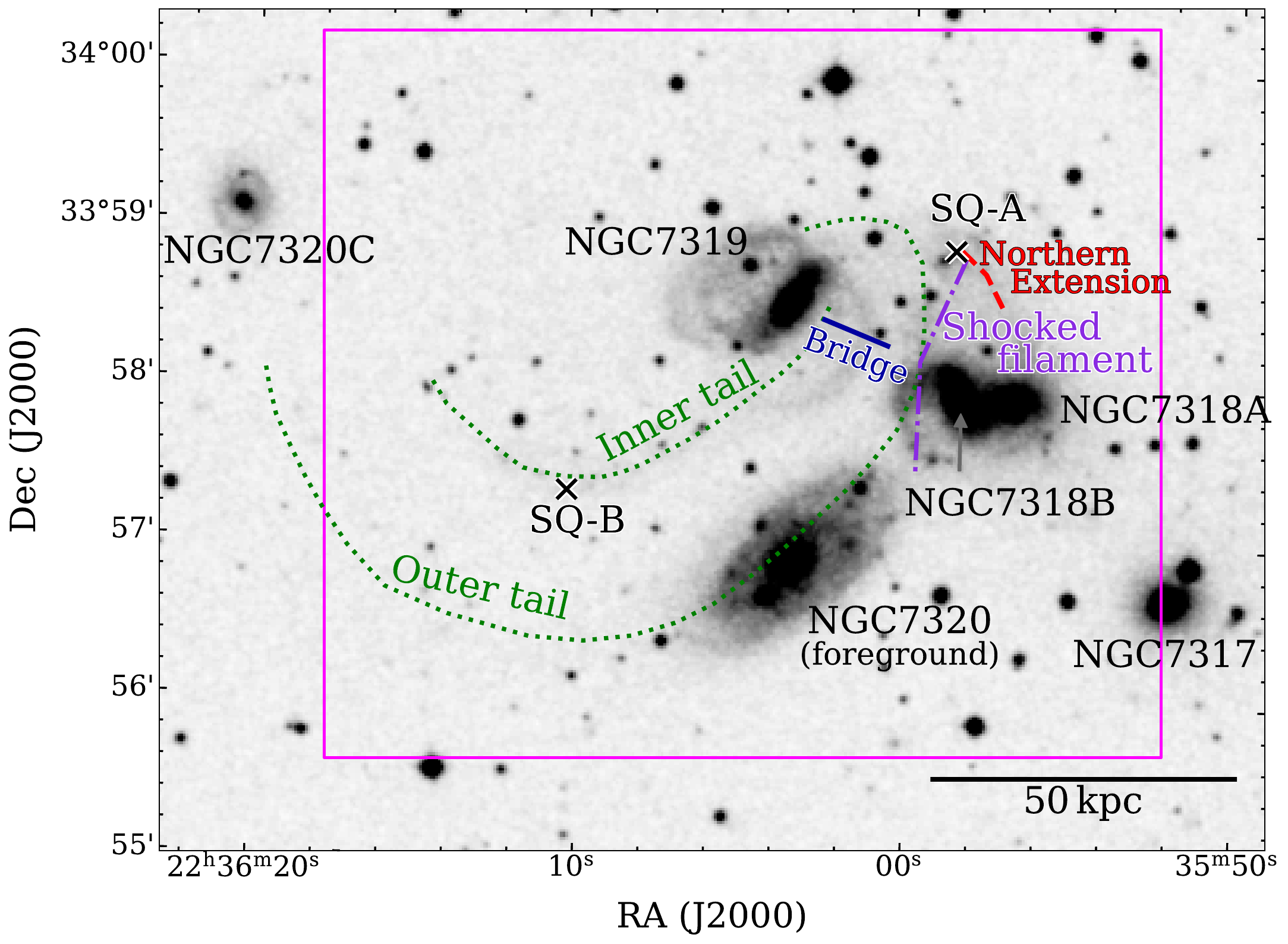} 
 \end{center}
\caption{
DSS2 $R-$band image of SQ with the galaxies in the field and characteristic structures labeled. The magenta rectangle indicates the ALMA ACA target field of view ($5^{\prime}.3 \times 4^{\prime}.6$).}\label{fig:SQ_guideline}
\end{figure}

Stephan's Quintet (SQ; Figure~\ref{fig:SQ_guideline}) is a well-known compact galaxy group.
It consists of five galaxies: NGC~7317, NGC~7318A/B, NGC~7319, and NGC~7320C, with NGC~7320 as the foreground galaxy.
VLA H\,{\sc i} mapping of SQ reveals that little H\,{\sc i} is present within the member galaxies, such as NGC~7319, while a large amount of H\,{\sc i} gas is distributed outside the galaxies, forming prominent tidal tails extending eastward from NGC~7319 \citep{Williams_2002AJ}.
This indicates that SQ is an evolved group in which H\,{\sc i} has been stripped from the member galaxies through multiple tidal interactions, resulting in H\,{\sc i} deficiency \citep[e.g.,][]{Verdes-Montenegro_2001A&A}.

SQ has several characteristic intergalactic morphological structures (see Figure~\ref{fig:SQ_guideline}), which reflect the significant and complex impact of repeated tidal interactions on intergalactic space in SQ.
Two prominent tidal tails (green dotted lines) are visible in deep optical images, and studies of star cluster candidates have shown that they have different ages, with the inner tail being younger (150–200~Myr) than the outer tail ($\sim$400~Myr) \citep{Fedotov2011AJ}.
In addition, a shocked filament (purple dash-dotted line) extends over $\sim$40~kpc in the north-south direction between NGC~7319 and NGC~7318B as seen in the JWST image \citep{Pontoppidan_2022ApJ}.
This structure is considered to be formed by the high-velocity ($\sim$900~$\mathrm{km~s^{-1}}$) intrusion of NGC~7318B into the group from the far side, colliding with the tidal debris produced by previous tidal interactions \citep[e.g.,][]{Sulentic_2001AJ,Xu_2003ApJ}.
Mid-infrared and H$\alpha$ observations \citep{Xu_1999ApJ} have confirmed active intergalactic star-forming regions (black cross marks): SQ-A, located to the north of NGC~7318A/B, and SQ-B, located within the inner tail.
Furthermore, warm H$_2$ observations have confirmed a bridge structure (navy line; hereafter referred to as Bridge) connecting the center of NGC~7319 and the shocked filament \citep{Cluver_2010ApJ}.

While numerous CO observations of SQ have been conducted in the past \citep[e.g.,][]{Yun_1997ApJ,Gao_Xu_2000ApJ,Smith_2001AJ,Lisenfeld_2002AA,Lisenfeld_2004AA,Petitpas_2005ApJ,Guillard_2012ApJ,Appleton_2023ApJ,Emonts_2025ApJ}, they have been limited to certain regions of the system. 
In particular, most of the observations in the tidal tail have been focused on the vicinity of SQ-B \citep{braine_abundant_2001,Lisenfeld_2002AA,Lisenfeld_2004AA}.
\citet{Emonts_2025ApJ} reported the large scale distribution of cold molecular gas with a spatial resolution of $\sim$3.2~kpc comprising NGC~7318A/B, shocked filament, SQ-A, and NGC~7319 using $^{12}$CO(2--1) data obtained with 7\,m array of the Atacama Compact Array (ACA) of the Atacama Large Millimeter/submillimeter Array (ALMA).
\citet{Maeda_2025ApJ} presented the first $^{12}$CO(1--0) mapping covering the full extent of the H\,{\sc i} distribution in SQ, including the tidal tail. 
Their wide-area CO(1--0) observations with the ACA Total Power (TP) array at a spatial resolution of $\sim$25\,kpc revealed molecular gas with low velocity dispersion ($\sim$20\,$\mathrm{km~s^{-1}}$) extending $\sim$100\,kpc along the inner tail and its northern area.

In this paper, we present a higher ($\sim$5.5\,kpc) resolution $^{12}$CO(1--0) map by adding the 7\,m array data to the TP data reported by \citet{Maeda_2025ApJ} and then investigate the spatial distribution and velocity structure of the molecular gas in the entire SQ system, including the tidal tail.
Furthermore, we aim to systematically clarify the differences in molecular gas mass and star formation activity among the main regions of the entire system.
We describe the observations and data reduction in Section~\ref{sec: observation}.
In Section~\ref{sec: results}, we present the spatial distribution and velocity structure of molecular gas.
Then, we compare the degree of star formation activity between the main structure of the entire SQ system, as well as investigate the relationship with the velocity dispersion of the CO(1--0) emission in Section~\ref{sec: discussion}. In this section, we also discuss the origin of the molecular gas in the tail region.
Finally, we summarize our main results and conclusions in Section~\ref{sec: conclusion}.
Throughout the paper, we assume a distance of 88.6~Mpc \citep{Fedotov_2015MNRAS,Duarte_Puertas_2019AA,Duarte_Puertas_2021AA}.

\section{Observations and Data Reduction}
\label{sec: observation}
The CO(1–0) mapping of the entire SQ was conducted with ALMA ACA (TP+7\,m array) in Band~3 under project 2023.1.01101.S (PI: F. Maeda). The field of view (FoV) spans $5^{\prime}.3 \times 4^{\prime}.6$ ($137\,\mathrm{kpc}\times119\,\mathrm{kpc}$; Figure~\ref{fig:SQ_guideline}), centered at ($22^\mathrm{h}36^\mathrm{m}05^\mathrm{s}.11$, $33^\circ57^\prime56^{\prime\prime}.7$) {(ICRS)},  covering all previously detected H\,{\sc i} emission \citep{Williams_2002AJ} and H\,{\sc ii} regions \citep{Duarte_Puertas_2019AA} in the system.
Details of the TP observations and their data reduction are presented in \citet{Maeda_2025ApJ}. In this section, we describe the 7\,m array observations and their reduction procedures.

The 7\,m array mosaic observations were carried out between December 16, 2023, and September 30, 2024, resulting in a total of 69 execution blocks (EBs). The average number of 7\,m antennas per EB was 9.6. The total number of mosaic fields is 46. The projected baseline length ranged from 8.9 m to 48.9 m, corresponding to a maximum recoverable scale of $\sim58^{\prime\prime}$ ($\sim$25\,kpc) at 115 GHz. The total on-source integration time is 87.6 hours.  The ACA correlator was used to obtain spectra with a bandwidth of 2.000~GHz and a channel spacing of 976.562~kHz, corresponding to $\sim$2.5~$\mathrm{km~s^{-1}}$. The spectral window was centered at 112.691~GHz, corresponding to the redshifted frequency of the CO(1--0) line at an optical velocity of 6861~$\mathrm{km~s^{-1}}$.
The typical system temperature for each EB ranged from 70 to 120~K.

We used the Common Astronomy Software Applications (CASA) package \citep{CASA_2022PASP} version 6.6.1.17 \citep[pipeline version 2024.1.0.8;][]{ALMA_pipeline_2023PASP} for the reduction of the 7\,m array data.
We applied the standard calibration scheme provided by the ALMA observatory. After concatenating all visibility files, we performed imaging.
We used the \texttt{tclean} task with the multi-scale deconvolver \citep{Cornwell_2008ISTSP} to recover extended emission as much as possible.
In the \texttt{tclean} task, we applied the Briggs weighting with a robust parameter of 0.5 in order to optimize both the sensitivity and angular resolution of the reduced image.
We used the \texttt{auto-multithresh} procedure \citep{Kepley_2020PASP} in \texttt{tclean} to identify automatically regions containing emission in the dirty and residual images.
We continued the deconvolution process until the peak intensity of the residual image reached the $\sim$1\,$\sigma$ noise level.

We combined the 7\,m and TP array data \citep{Maeda_2025ApJ} with the \texttt{feather} task to recover the total flux. The missing flux in the 7\,m-only data was 28$\%$.
The beam size of final CO data is 13$^{\prime\prime}$.6 $\times$ 12$^{\prime\prime}$.8 (5.85\,kpc $\times$ 5.48\,kpc), and the rms noise levels ($\sigma$) are 2.81\,mK (corresponding to $\sim$5.09\,Jy beam$^{-1}$) at a velocity resolution of 10\,km\,s$^{-1}$ and 2.06\,mK at a velocity resolution of 20\,km\,s$^{-1}$.
We primarily use the data with a velocity resolution of 20\,km\,s$^{-1}$.

\section{Results}
\label{sec: results}
\subsection{Global molecular gas properties}
\label{sec: global}

\begin{figure*}[p!]
 \begin{center}
  \includegraphics[width=180mm]{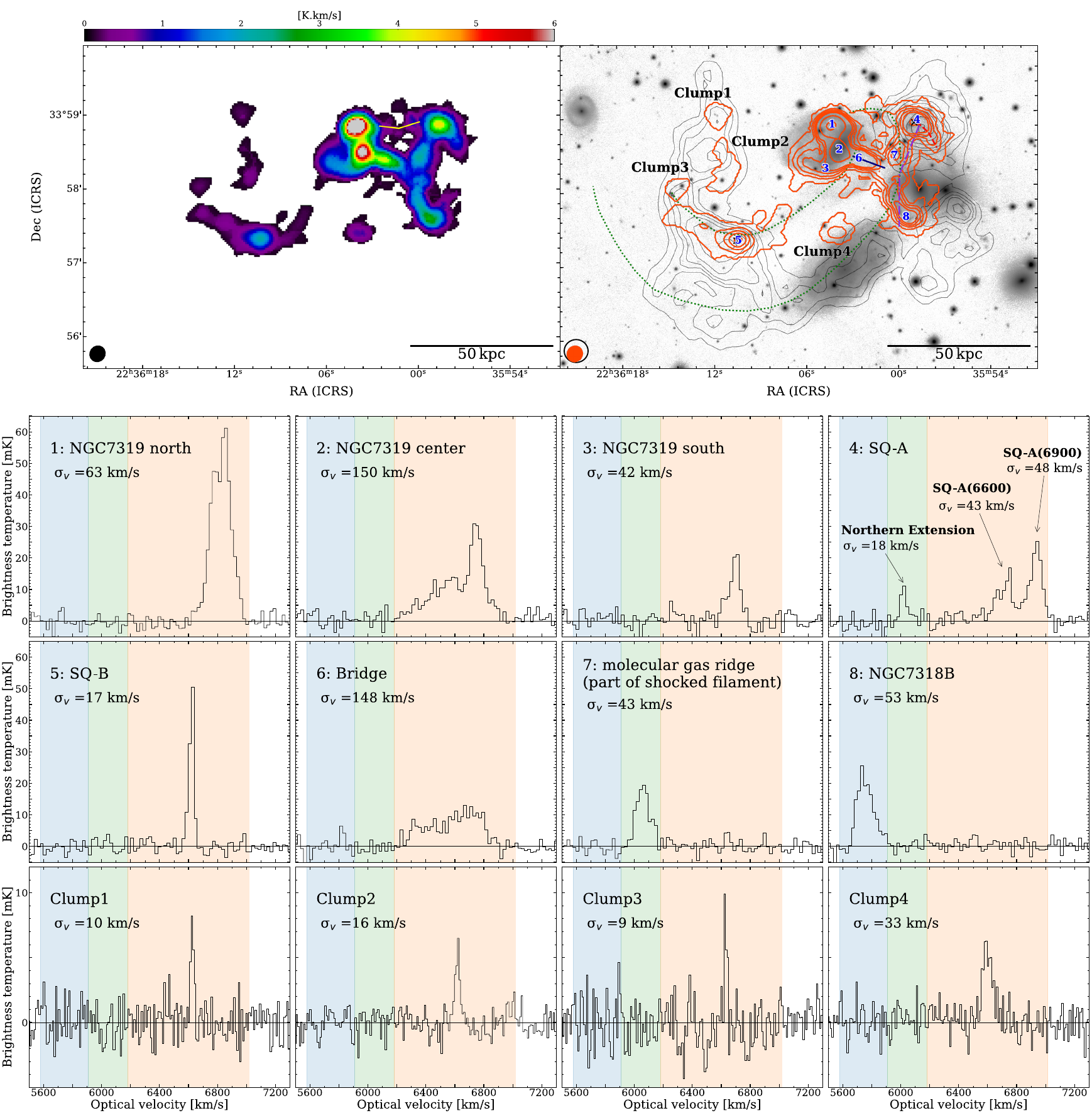} 
 \end{center}
\caption{
(Top left) Integrated intensity map of CO(1--0) in SQ over the entire velocity range from the masked cube. 
The yellow line shows the newly discovered bridge structure (see Section~\ref{sec: globalproperties}).
(Top right) Same CO(1--0) map (orange-red) overlaid on pan-STARRS $R\text{-}$band image with contour levels of mask boundary, $0.5, 1.0, 1.5, 2.0, 4.0$, and 8.0~K~$\mathrm{km\,s^{-1}}$. 
Black contours indicate H\,{\sc i} emission with a synthesized beam size of 19$^{\prime\prime}$.4 $\times$ 18$^{\prime\prime}$.6 \citep{Williams_2002AJ} at levels of  5.8, 15, 23, 32, 44, 61, 87, 120, and 180 $\times$ 10$^{19}$ atoms cm$^{-2}$.
The filled and open ellipses in the lower-left corner indicate the CO and H\,{\sc i} beams, respectively.
(Bottom) CO(1–0) spectra and their velocity dispersions ($\sigma_v$) derived from the spectra following the moment 2 definition at eight representative local peaks (marked by the blue numbers in the top-right panel) and region-averaged spectra of the CO clumps (Clump~1–4).
For CO clumps, cube data with a velocity resolution of 10~$\mathrm{km\,s^{-1}}$ were used instead of 20~$\mathrm{km\,s^{-1}}$.
The blue, green, and orange bands indicate the velocity ranges of three velocity components (see Section~\ref{sec: velocity components}).
The three peaks in the SQ-A spectrum corresponding to Northern Extension, SQ-A(6600), SQ-A(6900) (see Section~\ref{sec:SFEarea_define}) are shown with arrows.
}
\label{fig:SQmap+profile}
\end{figure*}

We defined significant emission using a mask that includes voxels with intensities above 2$\sigma$ adjacent to regions that satisfy either of the following criteria: (1) the signal exceeds 5$\sigma$, or (2) a signal of at least 4$\sigma$ is present across two adjacent channels.
The top panels in Figure~\ref{fig:SQmap+profile} display the CO(1–0) integrated intensity map of SQ over the full optical velocity range ($\sim$5580–7020\,km\,s$^{-1}$); the left panel shows the CO emission alone, while the right panel overlays the CO contours on the $R$-band image together with H\,{\sc i} contours \citep{Williams_2002AJ}.

CO(1--0) emissions were detected not only in the NGC~7319 disk, but also outside the galaxies, such as SQ-A, the shocked filament, and the inner tail. 
The existence of intergalactic molecular gas suggests that the SQ system is strongly affected by galaxy interactions. 
NGC~7319 contains a large amount of molecular gas in the galactic disk. 
However, there is little molecular gas on the disks of NGC~7318A/B except for the arm region of NGC~7318B. 
In the Bridge, we found spatially and kinetically continuous molecular gas connecting the shocked filament and the central region of NGC~7319 (distributed around the navy line in the top right panel of Figure~\ref{fig:SQmap+profile}). 
In addition, we have newly discovered another bridge structure connecting the northern region of NGC~7319 and SQ-A with a velocity of $\sim$6740~$\mathrm{km\,s^{-1}}$ (yellow line of the top left panel in Figure~\ref{fig:SQmap+profile}), which spans NGC~7319 and SQ-A.
The existence of this structure was also suggested in \citet{Maeda_2025ApJ}, but due to poor resolution, it was unclear whether it was truly a connected structure. 
Our finding of CO emission connecting the northern region of NGC~7319 and SQ-A implies that some of the gas in SQ-A has been stripped from NGC~7319, which is important for exploring the origin of the gas in SQ-A.
Neither bridge structure is clearly detected in the ACA 7\,m CO(2–1) map \citep{Emonts_2025ApJ}, although faint emission is reported in the Bridge.
The lack of clear detection likely reflects a difference in sensitivity or masking depth between CO(1-0) and CO(2-1) data sets, and/or extended CO(2–1) emission resolved out by the 7\,m array.

Our observation covers the entire H\,{\sc i} tidal tail extending to the east of NGC~7319 \citep{Williams_2002AJ}, including the region around SQ-B where CO had been detected in previous observations \citep{Lisenfeld_2002AA, Lisenfeld_2004AA}.
We discovered an elongated molecular gas structure including SQ-B and several molecular gas clumps (hereafter referred to as Clump~1--4; see the top panels of Figure~\ref{fig:SQmap+profile}) within the extended CO(1--0) emission in the tidal tail found by \citet{Maeda_2025ApJ}. 
The elongated molecular gas structure is present along the inner tail.
This molecular gas feature is more extended than that found by previous observations \citep{Lisenfeld_2002AA, Lisenfeld_2004AA}.
Three clumps (Clump~1--3) are located at the northeastern edge of the inner tail, and one clump (Clump~4) between the inner tail and the outer tail. 
However, no molecular gas structure was detected in the outer tail.

We estimate the total molecular gas mass of the entire masked regions in SQ (see the top panels of Figure~\ref{fig:SQmap+profile}) to be $(1.00\pm0.06)\times10^{10}~M_\odot$ under assumption of the Galactic CO-to-H$_2$ conversion factor of $\alpha_\mathrm{CO}=4.35~M_\odot~\mathrm{ (K~km\,s^{-1}~pc^2)^{-1}}$ \citep{bolatto_conversion_2013} considering that the metallicity of most H\,{\sc ii} regions in SQ, expressed as 12\,+\,log(O/H), are higher than 8.5 based on the O3N2 method \citep{Duarte_Puertas_2021AA}.
The uncertainty of the molecular gas mass includes the absolute flux calibration error in ALMA data (5$\%$) and the error in measuring the CO intensity due to its rms noise.
The total molecular gas mass is consistent with that observed by ACA TP, $(1.07\pm0.05)\times10^{10}~M_\odot$ \citep{Maeda_2025ApJ}, within the errors.
The molecular gas masses and their uncertainties presented below are calculated under the same assumptions and method.

Here, we compare the spatial distribution of CO and H\,{\sc i}. 
Note that this comparison is still tentative because the velocity range of H\,{\sc i} data by \citet{Williams_2002AJ} is 5725--6918~$\mathrm{km\,s^{-1}}$ in optical definition, which does not fully cover the range in which significant CO(1--0) emission is detected.
A quantitative comparison between CO and H\,{\sc i} will be the subject of a future work.
In NGC~7319 and the eastern spiral arm of NGC~7318B, a substantial amount of molecular gas is detected, but little H\,{\sc i} gas is observed.
This suggests that past galaxy interactions may have preferentially stripped H\,{\sc i} gas, while molecular gas has largely remained in place.
Molecular gas in the tidal tail is mainly associated with H\,{\sc i} gas, but the part near NGC~7319 (i.e., Clump~4 and the western side of SQ-B) is not associated with H\,{\sc i} gas (see Section~\ref{sec: tails}).
There is a lot of molecular gas and H\,{\sc i} gas around SQ-A.
No notable molecular gas corresponding to the H\,{\sc i} emission located south of NGC~7318A has been found.
In this region, significant CO emission has not been detected in previous observations (CO(1--0); \citet{Lisenfeld_2002AA}, CO(2--1); \citet{Emonts_2025ApJ}), whereas star-forming regions have been confirmed \citep[e.g.,][]{Natale_2010ApJ, Duarte_Puertas_2021AA}.
\label{sec: globalproperties}

\begin{figure}[t!]
 \begin{center}
  \includegraphics[width=85mm]{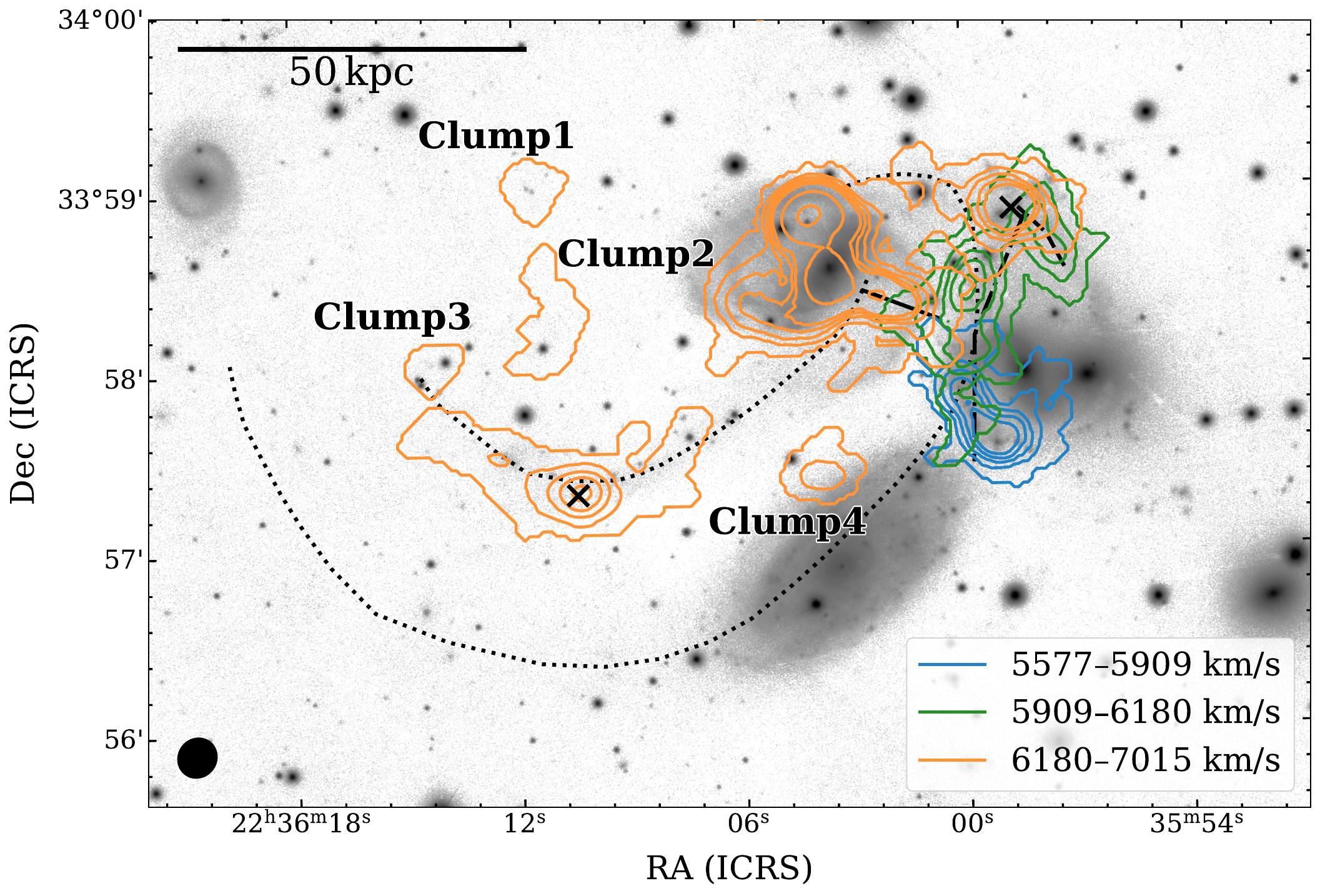} 
 \end{center}
\caption{
Integrated intensity map divided into three velocity components (blue: low-, green: mid-, orange: high-velocity component).
The contour levels are the same as in Figure~\ref{fig:SQmap+profile}.
See Section~\ref{sec: velocity components} for details.} \label{fig:mom0each}
\end{figure}

\subsection{CO velocity structures}
\label{sec: velocity components}

The molecular gas in SQ is distributed over a wide velocity range up to $\sim$1400~$\mathrm{km\,s^{-1}}$.
The bottom panel of Figure~\ref{fig:SQmap+profile} shows the spectra of eight representative positions and the region-averaged spectra, which are integrated over the area enclosed by the mask boundary, for four CO clumps.
They reveal that the velocity components significantly vary depending on the regions as reported in previous studies \citep[e.g.,][]{Lisenfeld_2002AA,Guillard_2012ApJ,Emonts_2025ApJ, Maeda_2025ApJ}.
In this subsection, we divide CO emission into three velocity components and discuss each of them in detail.
The velocity ranges for the low-, mid-, and high-velocity components are defined as 5577--5909~$\mathrm{km\,s^{-1}}$, 5909--6180~$\mathrm{km\,s^{-1}}$, and 6180--7015~$\mathrm{km\,s^{-1}}$, respectively.
While the three velocity components are classified largely according to \citet{Maeda_2025ApJ}, the boundary between the mid- and high-velocity ranges is adjusted to $6180\,\mathrm{km\,s^{-1}}$. 
This boundary corresponds to the velocity transition between the significant emissions of the shocked filament and Bridge, as illustrated by a comparison of their line profiles in Figure~\ref{fig:SQmap+profile}. 
Figure~\ref{fig:mom0each} shows the integrated intensity maps separated into three velocity components.

Low-velocity component (5577--5909\,$\mathrm{km\,s^{-1}}$):
This velocity component is associated with NGC~7318B (systemic velocity: 5765~$\mathrm{km\,s^{-1}}$; \citealt{Sulentic_2001AJ}), and is distributed primarily along the eastern spiral arm of NGC~7318B.
CO intensity is the strongest in the southern part of the arm.
This region is an active star-forming site located at the southern area of the shocked filament \citep[e.g.,][]{Cluver_2010ApJ}.
The molecular gas mass of this component is $(1.0\pm0.1)\times10^{9}~M_\odot$, which accounts for about 10\% of the total molecular gas mass in SQ.

Mid-velocity component (5909--6180\,$\mathrm{km\,s^{-1}}$):
This velocity component kinematically connects the molecular gas associated with NGC~7318B and NGC~7319.
CO emission is mostly detected outside of galaxies; One is Northern Extension, which corresponds to one of the several velocity components in SQ-A (see the spectrum of Figure~\ref{fig:SQmap+profile} and also \citet{Emonts_2025ApJ}).
In addition, a molecular gas ridge runs along the shocked filament between the SQ-A and the eastern spiral arm of NGC~7318B.
Our CO(1--0) map reveals spatially and kinematically continuous emission bridging the molecular gas ridge and the Northern Extension.
While the CO(2–1) observations \citep{Emonts_2025ApJ} show a clear spatial separation between the shocked filament and the Northern Extension, our CO(1–0) observations reveal spatially and kinematically continuous emission bridging the two.
This suggests that the regions are physically connected by spatially extended molecular gas that is faint or undetected in CO(2-1).
The molecular gas mass of this velocity component is $(1.4\pm0.1)\times10^{9}~M_\odot$, which accounts for about 14\% of the total molecular gas mass in SQ.

High-velocity component (6180--7015\,$\mathrm{km\,s^{-1}}$):
This velocity component contains molecular gas spread over a wide area in SQ, not only associated with NGC~7319 but also located in the Bridge, tidal tail (including SQ-B), and SQ-A.
In the central and northern parts of NGC~7319, two velocity components are confirmed in each CO spectrum (see Figure~\ref{fig:SQmap+profile}), consistent with the ACA CO(2--1) observations \citep{Emonts_2025ApJ}.
The molecular gas components in the tidal tail (SQ-B, Clump~1--4) have very narrow linewidths compared to other regions in SQ, with the strongest concentration in SQ-B.
In contrast, the broad component is detected in the Bridge, as also reported in previous studies \citep{Guillard_2012ApJ,Emonts_2025ApJ}.
The total molecular gas mass of the high-velocity component is ($7.6\pm0.5)$$\times$10$^{9}$ $M_\odot$, which accounts for about 76\% of the total molecular gas mass in SQ.

\subsection{Molecular gas structure along the inner tail and CO clumps}
\label{sec: clumps}
Previous CO observations of the tidal tail \citep[e.g.,][]{braine_abundant_2001,Lisenfeld_2002AA,Lisenfeld_2004AA} were limited to the vicinity of the star-forming region SQ-B and to a star-forming region located east of SQ-B (known as SQ tip; \citet{Lisenfeld_2004AA}).
In contrast, our wide-area CO mapping reveals molecular gas extending along the optically identified inner tidal tail, including SQ-B and four discrete CO clumps (Clump 1–4; Section~\ref{sec: global}, Figure~\ref{fig:SQmap+profile}).
The basic properties of the detected molecular gas structures are summarized in Table~\ref{tab:clumps}.
Center coordinates (Columns~2--3), deconvolved major/minor axis (Column~4), and position angle (Column~5) were derived from two-dimensional Gaussian fitting to the integrated intensity map.
Note that no Gaussian fitting was performed for SQ-B due to its prominently elongated morphology; the listed position corresponds to the peak of CO intensity.
The total molecular gas masses (Column~6) were obtained from the integrated intensity map.
Radial velocity (Column~7) and velocity dispersion (Column~8) were derived from the Gaussian fitting of the averaged CO spectra at a velocity resolution of 10~$\mathrm{km\,s^{-1}}$.

The total mass of molecular gas of SQ-B, including an extending component in the east-west direction from SQ-B, is estimated to be $(75.5\pm5.5)\times10^{7}~M_\odot$, and the velocity dispersion is about 19~$\mathrm{km\,s^{-1}}$.
The molecular gas masses of three clumps (Clump~1--3) are in the order of 10$^{7}$~$M_\odot$ and the velocity dispersions are less than 20~$\mathrm{km\,s^{-1}}$.
However, Clump~4 is more massive ($>10^8 \, M_\odot$) and shows slightly larger velocity dispersion ($33~\mathrm{km \, s^{-1}}$) compared to Clump~1--3.
The origins of Clump~1--4 are discussed in Section~\ref{sec: tails}.

The star formation activity and velocity dispersion significantly vary across different regions, even within the same velocity component. We discuss these properties in detail in Section~\ref{sec:SFE_vs_sigma}.

\begin{deluxetable*}{cccccccc}
\tablewidth{0pt} 
\tablecaption{Physical properties of molecular gas structures in the tail region
\label{tab:clumps}}
\tablehead{
\colhead{Name}  & \colhead{R.A.} & \colhead{Decl.} & \colhead{$\theta_{\rm maj} \times \theta_{\rm min}$}  & P.A.          & \colhead{Mass}      & \colhead{$v_{\rm rad}$} & \colhead{$\sigma_v$}   \\[-1.0mm]
                & (hh:mm:ss.ss)  & (dd:mm:ss.s)  & (arcsec$^2$)                                          & (deg.)        & ($10^7\,M_{\odot}$) & (km\,s$^{-1}$)          & (km\,s$^{-1}$)      \\[-0.5mm]
\colhead{(1)}   & \colhead{(2)}  & \colhead{(3)}   & \colhead{(4)}       & \colhead{(5)} & \colhead{(6)}    & \colhead{(7)}   & \colhead{(8)}
} 
\startdata 
*SQ-B      & 22:36:10.30    & 33:57:19.1   & --		             & --      & 75.5 $\pm$ 5.5   &  6620 $\pm$ 4   & 18.5 $\pm$  0.6  \\
Clump~1	  & 22:36:11.54    & 33:59:01.1   & 20.0 $\times$ 17.4   & 13.1    & 3.2 $\pm$ 0.9     &  6625 $\pm$ 4   & 10.5 $\pm$  1.8  \\
Clump~2	  & 22:36:10.99    & 33:58:10.6   & 46.8 $\times$ 16.3   & -0.8    & 7.5 $\pm$ 1.8     &  6619 $\pm$ 5   & 15.7 $\pm$  1.9  \\
Clump~3	  & 22:36:14.27    & 33:58:02.4   & 21.4 $\times$ 14.5 	 & -60.0   & 2.7 $\pm$ 0.6     &  6628 $\pm$ 5   &  9.3 $\pm$  2.0  \\
Clump~4	  & 22:36:03.85    & 33:57:24.5   & 21.5 $\times$ 11.8	 & 81.0    & 11.5 $\pm$ 1.8    &  6598 $\pm$ 5   & 33.3 $\pm$  3.6  \\
\enddata          
\tablecomments{
(1) Name of the structure.
(2) -- (3) CO($J=1-0$) peak position of the clump in equatorial coordinates (ICRS).
(4) Deconvolved major/minor axis of the clump in units of arcsec.
(5) Position angle of the deconvolved major axis of the clump, which is measured counterclockwise from north to east, in degree.
(6) Molecular gas mass of the clump derived from the CO($J=1-0$) line intensity under the assumption of the Galactic $\alpha_\mathrm{CO}$ of 4.35~$M_\odot~\mathrm{ (K~km\,s^{-1}~pc^2)^{-1}}$ including uncertainty in units of $10^7\,M_{\odot}$.
(7) Radial velocity including uncertainty in units of km\,s$^{-1}$ (LSR, optical definition).
(8) Velocity dispersion of the clump including uncertainty in units of km\,s$^{-1}$.\\
(*) The molecular gas mass, radial velocity, and velocity dispersion of SQ-B are derived including the east-west extending component.
}
\end{deluxetable*}

\section{Discussions}
\label{sec: discussion}
Using our CO mapping data that covers the entire SQ system, including the tidal tails, we investigate the star formation efficiency (SFE, defined as the ratio of star formation rate (SFR) to molecular gas mass) in each region to perform a relative comparison of star formation activities across the major structures in SQ.

\subsection{Variation in star formation activity among major structures}
\label{sec:SFEarea_define}
In this subsection, we describe the SFE estimation procedure and the resulting differences between major structures.

\subsubsection{Definition of molecular gas regions}
Our CO map revealed that the cold molecular gas in SQ is widely distributed over intergalactic regions as well as in the main bodies of member galaxies such as NGC~7319.
This requires us to choose appropriate ``molecular gas regions'' for the SFE calculation.
Here, we describe the objective criteria to define the molecular gas regions in SQ as follows.

First, we identify peaks in the integrated intensity map of each velocity component defined in Section~\ref{sec: velocity components} and define regions by expanding from each peak to include adjacent pixels with intensities exceeding 20\% of the peak value.
Note that this pixel expansion terminates at the mask boundary defined in Section~\ref{sec: global} even if the CO intensity is still above 20\% of the peak value.
When two neighboring peaks are separated by more than one beam and fall within the same region, the locus of lowest intensity (i.e., the intensity valley) in the integrated intensity map between the peaks is adopted as the boundary.

\begin{figure}[t!]
 \begin{center}
  \includegraphics[width=85mm]{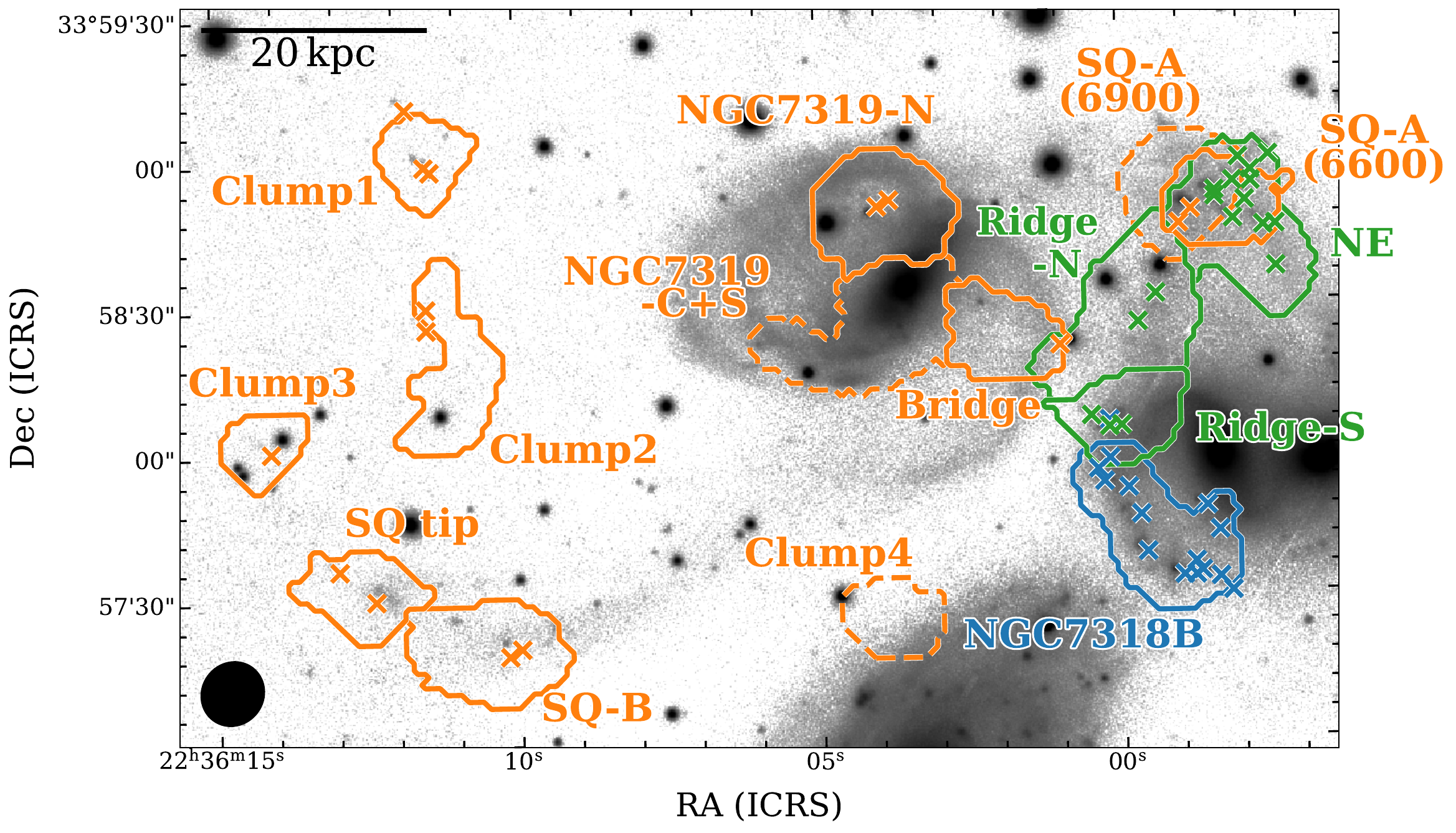} 
 \end{center}
\caption{
Molecular gas regions (contours) and associated H\,{\sc ii} regions (cross marks) used to derive the SFE.
The contour colors indicate velocity components (blue: low-, green: mid-, orange: high-velocity component). 
Ridge north, Ridge south, and Northern Extension are abbreviated as Ridge-N, Ridge-S, and NE, respectively. 
SFRs and SFEs are not derived in the regions indicated by dashed lines (NGC~7319-C+S, SQ-A(6900), and Clump~4).}\label{fig:SFEarea}
\end{figure}

Figure~\ref{fig:SFEarea} shows the determined molecular gas regions in SQ.
In most cases, the selected regions correspond to the representative positions in which the CO spectra are presented in Figure~\ref{fig:SQmap+profile}. However, the region definitions become more complex in areas such as the molecular gas ridge, NGC~7319, SQ-A, and SQ-B.

Based on our region‐division criteria, the molecular gas ridge of the mid‐velocity component is divided into two regions, hereafter designated as Ridge-N (Ridge north) and Ridge-S (Ridge south).
On the other hand, the central and southern regions of NGC~7319 cannot be separated. This is because CO intensities are seamlessly connected between the two regions without any intensity valley. 
Thus, we treated these two regions as a single region (NGC~7319-C+S).

\begin{figure}[t!]
 \begin{center}
  \includegraphics[width=85mm]{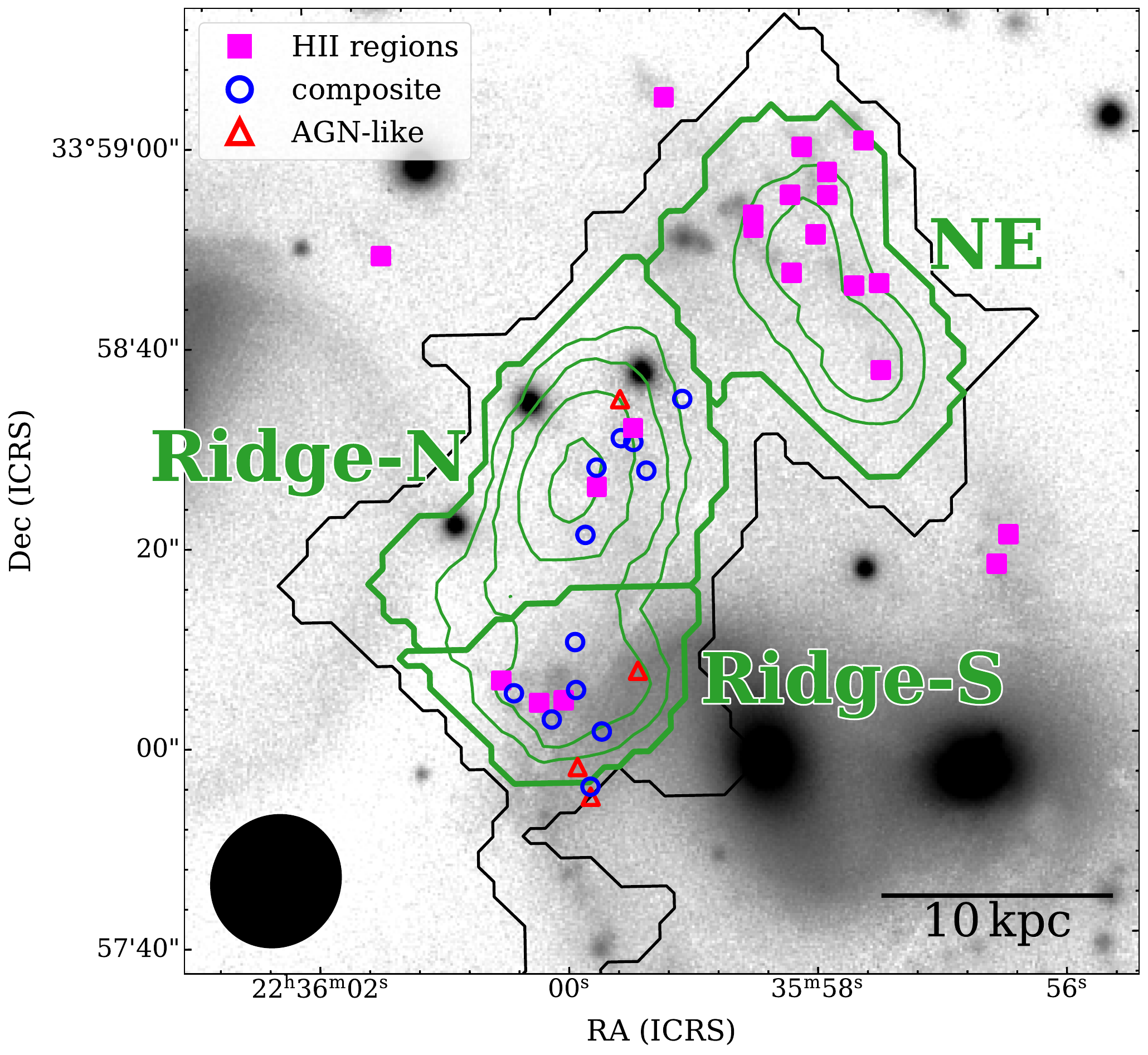} 
 \end{center}
\caption{
Positions and classifications based on the BPT diagram for H$\alpha$ emitters in the mid-velocity component. 
The black contour shows the boundary of the CO mid-velocity component.
The green contours indicate the regions used for the SFE calculation (thick) and integrated intensity levels of 0.7, 1.0, 1.5, and 2.0~K~$\mathrm{km\,s^{-1}}$.
The H$\alpha$ emitters classified as AGN-like or composite are likely affected by shock \citep{Duarte_Puertas_2021AA} and not included in the SFE calculation.}\label{fig:ShBPT}
\end{figure}

In the high-velocity component of SQ-A, two subcomponents are present (see the spectrum of Figure~\ref{fig:SQmap+profile}). We therefore separated the integrated intensity map into two velocity intervals and treated these subcomponents as independent regions, named SQ-A(6600) and SQ-A(6900), using a velocity boundary of 6810~$\mathrm{km\,s^{-1}}$, where the intensity between them reaches a minimum.
Although Northern Extension (hereafter referred to as NE) has two CO peaks, it was defined as a single region because their separation is smaller than the beam size (see Figure~\ref{fig:ShBPT}).
The regions labeled Clump~1 to Clump~4 correspond to the CO clumps defined in Section~\ref{sec: global}.\footnote{According to our region-division criteria, Clump~2 and Clump~4 are a little bit smaller than those defined by the mask boundary shown in Figure~\ref{fig:SQmap+profile}. Nevertheless, the estimated molecular gas masses are almost consistent between the two definitions within the errors (see Tables~\ref{tab:clumps} and \ref{tab:regions}). Thus, we also use Clump~2 and Clump~4 as the region names in the following sections.}
The SQ tip region named by \citet{Lisenfeld_2004AA} is separated from SQ-B.

We derived molecular gas mass by summing up the CO intensity included in each molecular gas region.
Note that the extent of the molecular gas region and the resulting molecular gas mass vary depending on the adopted lowest intensity level relative to the peak value.
We calculated molecular gas masses in two ways, adopting the lowest intensity levels of 10\% and 30\% in each region, and treated them as the uncertainty range.
Combining this uncertainty, the rms error, and the absolute flux calibration error, we estimated the final error of molecular gas mass.

\subsubsection{Deriving the SFRs and the SFEs}
\label{sec:SFR_SFE}
We use the SFRs derived from H$\alpha$ luminosities of cataloged star-forming H\,{\sc ii} regions, which are corrected for the dust extinction using the Balmer decrement \citep{Duarte_Puertas_2019AA,Duarte_Puertas_2021AA}. 
Although star formation can also be traced by infrared and/or ultraviolet emission, the H$\alpha$-based catalog using an imaging Fourier transform spectrometer allows individual star-forming regions to be separated in velocity space, which is essential for systems like SQ where multiple velocity components overlap along the line of sight.
In addition, the use of the Baldwin-Phillips-Terlevich (BPT) diagram enables the identification of H$\alpha$ emitters dominated by star formation, minimizing contamination from shocks or active galactic nuclei \citep[AGN;][]{Duarte_Puertas_2021AA}.

In deriving the SFR, we included all H\,{\sc ii} regions that satisfy both of the following criteria: (1) located within half a beam from each region boundary, and (2) the H$\alpha$ radial velocity falls within the CO velocity range of the associated molecular gas region.
However, we cannot determine the SFRs for some molecular gas regions by this method due to the following reasons.

Although the central region of NGC~7319 has extremely bright H$\alpha$ emitters, almost all of them are classified as AGN-type by the BPT diagram.
This makes it difficult to estimate the $true$ SFR in this region, even if the nuclear star formation coexists, due to quite large uncertainty.
Meanwhile, the southern region of NGC~7319 includes two H$\alpha$ emitters, yet both of them are unclassified in the BPT diagram because of the lack of essential optical emission-line data.
In this case, H$\alpha$ luminosities of the emitters are not measured, which makes it impossible to give any constraint on the SFR.
A similar situation applies to SQ-A(6900); this region also includes several highly luminous H$\alpha$ emitters, but they are unclassified in the BPT diagram for the same reason.
Although \citet{Xu+25} suggested the suppression of star formation in SQ-A(6900) based on their JWST 15~$\mu$m observations, the SFR is still highly uncertain.
Therefore, we do not derive the SFRs in NGC~7319-C+S and SQ-A(6900).

Interestingly, no H$\alpha$ emitter is found in Clump~4 in spite of its large molecular gas reservoir ($\sim$\,$10^8\,M_{\odot}$).
In addition, Clump~4 does not have any counterparts at other wavelengths such as infrared, indicating the absence of dust-obscured star formation and truly low SFE.
We do not quantify the SFR in Clump~4, but instead we will discuss the nature of Clump~4 in detail in Section~\ref{sec: tails}.

Eventually, we derived SFRs and SFEs for 12 molecular gas regions in SQ, as shown in Table~\ref{tab:regions}.

\begin{figure*}
 \begin{center}
  \includegraphics[width=180mm]{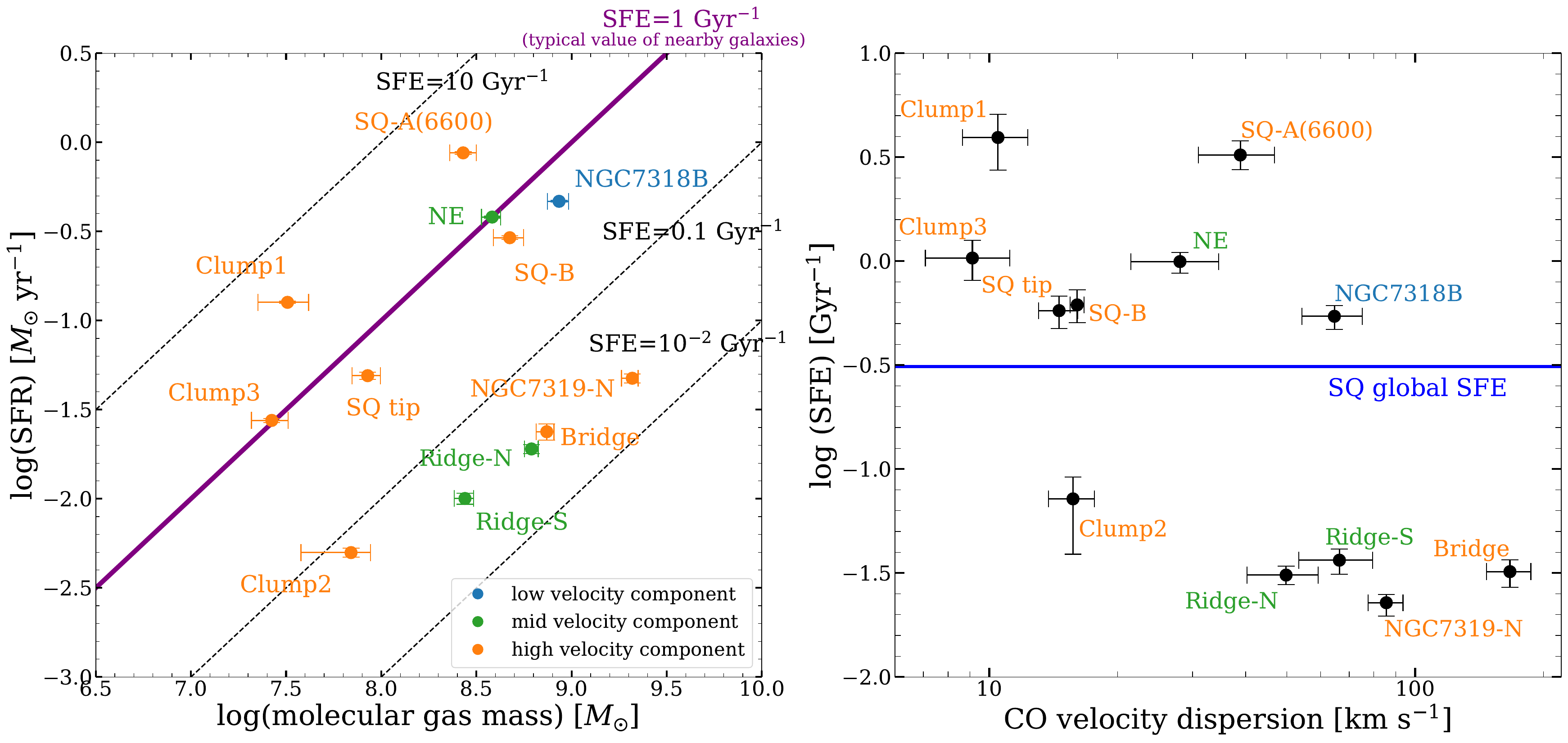} 
 \end{center}
\caption{
(Left) Relation between molecular gas mass and SFR for each region in SQ. 
The region names and colors are the same as Figure~\ref{fig:SFEarea}. 
The purple line shows the SFE of 1\,Gyr$^{-1}$, a typical value in nearby star forming galaxies \citep{Leroy2008AJ,Muraoka_2019PASJ}. 
(Right) Correlation between SFE and CO velocity dispersion. 
Global SFE in SQ ($\sim$0.31~Gyr$^{-1}$) is indicated in blue.
}
\label{fig:KSlaw+SFE_sigma}
\end{figure*}

\begin{deluxetable*}{cccc}
\tablewidth{0pt} 
\tablecaption{Molecular gas mass, SFRs, and SFEs in molecular gas regions
\label{tab:regions}}
\tablehead{
\colhead{Region name}  & \colhead{log $M_{\rm mol}$}  & \colhead{log SFR}            & \colhead{log SFE} \\[-1.0mm]
\colhead{}             & \colhead{($M_{\odot}$)} & \colhead{($M_{\odot}$~yr$^{-1}$)}  & \colhead{(Gyr$^{-1}$)}
}
\startdata
NGC~7318B     & $8.93^{+0.05}_{-0.06}$    & $-0.33^{+0.01}_{-0.01}$    & $-0.27^{+0.05}_{-0.06}$   \\ [0.5mm]
NE            & $8.58^{+0.04}_{-0.06}$    & $-0.42^{+0.01}_{-0.01}$    & $0.00^{+0.04}_{-0.06}$    \\ [0.5mm]
Ridge-N       & $8.79^{+0.04}_{-0.04}$    & $-1.72^{+0.02}_{-0.03}$    & $-1.51^{+0.04}_{-0.05}$   \\ [0.5mm]
Ridge-S       & $8.44^{+0.05}_{-0.06}$    & $-2.00^{+0.03}_{-0.03}$    & $-1.44^{+0.05}_{-0.07}$   \\ [0.5mm]
Bridge        & $8.87^{+0.04}_{-0.06}$    & $-1.62^{+0.04}_{-0.05}$    & $-1.49^{+0.06}_{-0.08}$   \\ [0.5mm]
NGC~7319-N    & $9.32^{+0.03}_{-0.06}$    & $-1.32^{+0.03}_{-0.03}$    & $-1.64^{+0.04}_{-0.06}$   \\ [0.5mm]
NGC~7319-C+S  & $9.26^{+0.07}_{-0.11}$    &  ---                       & ---                       \\ [0.5mm]
Clump~1       & $7.51^{+0.11}_{-0.16}$    & $-0.90^{+0.01}_{-0.01}$    & $0.59^{+0.11}_{-0.16}$    \\ [0.5mm]
Clump~2       & $7.84^{+0.10}_{-0.26}$    & $-2.30^{+0.02}_{-0.03}$    & $-1.14^{+0.10}_{-0.27}$   \\ [0.5mm]
Clump~3       & $7.42^{+0.09}_{-0.11}$    & $-1.56^{+0.01}_{-0.01}$    & $0.02^{+0.09}_{-0.11}$    \\ [0.5mm]
Clump~4       & $8.02^{+0.07}_{-0.07}$    &  ---                       & ---                       \\ [0.5mm]
SQ tip        & $7.93^{+0.07}_{-0.08}$    & $-1.31^{+0.02}_{-0.02}$    & $-0.24^{+0.07}_{-0.09}$   \\ [0.5mm]
SQ-B          & $8.68^{+0.07}_{-0.08}$    & $-0.54^{+0.01}_{-0.01}$    & $-0.21^{+0.07}_{-0.09}$   \\ [0.5mm]
SQ-A(6600)    & $8.43^{+0.07}_{-0.07}$    & $-0.06^{+0.01}_{-0.01}$    & $0.51^{+0.07}_{-0.07}$    \\ [0.5mm]
SQ-A(6900)    & $8.74^{+0.06}_{-0.07}$    &  ---                       &  ---                      \\ [0.5mm]
\enddata          
\tablecomments{
$M_{\rm mol}$ means the molecular gas mass calculated from the total CO(1--0) luminosity within each molecular gas region shown in Figure~\ref{fig:SFEarea}, assuming the Galactic $\alpha_\mathrm{CO}$ of 4.35~$M_\odot~\mathrm{ (K~km\,s^{-1}~pc^2)^{-1}}$.
}
\end{deluxetable*}

\subsubsection{Comparison of SFE for each region in SQ}
\label{sec:compareSFE}
The left panel of Figure~\ref{fig:KSlaw+SFE_sigma} shows a diversity of SFEs over two orders of magnitude among regions.
In many regions, including intergalactic ones like SQ-A(6600), SQ-B, Clump~1, and Clump~3, the SFEs fall within 0.6--4.0\,Gyr$^{-1}$, comparable to that of nearby star-forming galaxies \citep[$\sim$1\,Gyr$^{-1}$;][]{Leroy2008AJ,Muraoka_2019PASJ} (hereafter moderate SFE regions).
In contrast, the SFEs in NGC~7319-N, Ridge-N, Ridge-S, Bridge, and Clump~2 are at least ten times lower than in nearby galaxies (0.023--0.072\,Gyr$^{-1}$, hereafter low SFE regions).

Based on a comparison between previous studies and their own observations, \citet{Maeda_2024} suggested that SFE values in tidal tails span a wide range from $10^{-2}$ to $10\text{~Gyr}^{-1}$.
Our measurements in the most of the tail regions (Clump~1, Clump~3, SQ-B, and SQ tip) range from 0.6 to $4.0\text{~Gyr}^{-1}$, placing them within
the relatively high SFE regime of this distribution.
In particular, Clump~1 exhibits the highest SFE; this is consistent with the recent active star formation in this region, as evidenced by the presence of blue star clusters \citep{Gallagher_2001AJ} and the young age of the H\,{\sc ii} regions (a few Myr) associated with these clusters \citep{Mendes_de_Oliveira_2004ogci}.
However, the remaining region (Clump~2) has a low SFE ($\sim$0.072\,Gyr$^{-1}$).
We discuss the difference in SFE among CO clumps in Section~\ref{sec: tails}.

NGC~7319-N has a large amount of molecular gas ($\sim$2.1$\times$10$^{9}$~$M_\odot$), but a low SFR ($\sim$0.05~$M_\odot$~yr$^{-1}$).
A recent study of star cluster candidates based on HST and JWST imaging \citep{Aromal_2025ApJ} found that NGC\,7319 lacks young star clusters. This may be related to the suppression of star formation in NGC\,7319-N. However, NGC\,7319-N lies outside the area covered by the HST imaging, and therefore no star cluster candidates of any age have been identified in this region.

In the Ridge, earlier studies have also suggested that star formation is suppressed \citep[e.g.,][]{Cluver_2010ApJ,Guillard_2012ApJ}.
This region has been confirmed to be affected by the shock resulting from the ongoing collision between NGC~7318B and the intergalactic medium through optical \citep{Xu_2003ApJ} and mid infrared \citep{Cluver_2010ApJ} emission line diagnostics.
Moreover, only a few H$\alpha$ emitters are classified as H\,{\sc ii} regions based on the BPT diagram \citep{Duarte_Puertas_2021AA}.
Figure~\ref{fig:ShBPT} shows the positions of H$\alpha$ emitters whose velocities fall within the mid-velocity component and their classification based on the BPT diagram.
As seen in this figure and reflected in the SFE measurements, the star formation activity significantly differs between NE and the Ridge-N/S.
The Ridge contains some H$\alpha$ emitters classified as composite or AGN-like regimes, indicating that shocks have a strong influence on the intra-group medium (IGM).
In fact, a detailed analysis of optical emission lines in SQ by \citet{Konstantopoulos_2014ApJ} found that some H$\alpha$ emitters in the Ridge can be explained by the shock model. 
These emitters are classified as composite or AGN-like in \citet{Duarte_Puertas_2021AA}, supporting the interpretation that shocks yield such classifications.

The Bridge is almost devoid of star-forming regions \citep{Duarte_Puertas_2021AA}.
This region exhibits the largest velocity dispersion among the SQ regions (167~$\mathrm{km~s^{-1}}$), which may be related to the suppression of star formation. 
A detailed discussion is presented in Section~\ref{sec:SFE_vs_sigma}.

\subsection{Relation between SFE and velocity dispersion}
\label{sec:SFE_vs_sigma}
As mentioned in the previous subsection, we found variation in SFE within SQ.
One possible cause of this is gas turbulence.
In this subsection, we focus on CO velocity dispersion in each region as a proxy of turbulence.
The velocity dispersions were calculated based on the intensity-weighted second moment, using only contiguous channels that have significant CO emission within each region.
We used the cube data with a velocity resolution of 10~$\mathrm{km~s^{-1}}$ for the tail regions where the CO lines are narrow (Clump~1--3, SQ-B, and SQ tip), while using that of 20~$\mathrm{km~s^{-1}}$ for other regions.

The right panel of Figure~\ref{fig:KSlaw+SFE_sigma} plots SFE against CO velocity dispersion, showing a clear anti-correlation with a Spearman's rank correlation coefficient of -0.72.
A similar correlation is found in the bar and bar-end regions of the gas-rich long-bar galaxies \citep{Maeda_2023ApJ}.
The velocity dispersions in low SFE regions tend to be larger than those in moderate SFE regions.

The trend of large velocity dispersions in low SFE regions suggests that the suppression of star formation is linked to the turbulent motions.
Several physical interpretations can be considered.
One possibility is that large-scale turbulence inhibits the formation of giant molecular clouds (GMCs), which results in the low efficiency of high-mass star formation.
Alternatively, GMCs are present, but their velocity dispersions could be large enough to prevent gravitational collapse and star formation.
Besides, it is also possible that high relative motions among the clouds cause high-speed cloud-cloud collisions, leading to the suppression of star formation, as suggested by numerical/observational studies of the bar regions of barred galaxies \citep{Fujimoto_2014MNRAS,Maeda_2020MNRAS_GMC_properties,Maeda_2020MNRAS_diffuse_gas_SFE,Maeda_2025ApJ_3627}.

In the Bridge, the velocity dispersion is significantly larger than in other regions.
ALMA 12\,m array observations of a part of the Bridge region at $\sim$120\,pc resolution \citep{Appleton_2023ApJ} revealed that the velocity dispersions of CO emission within individual beam-sized apertures are notably large (FWHM$\sim$100~$\mathrm{km~s^{-1}}$) and highly turbulent, which may prevent gravitational collapse and star formation.
The authors also reported that those clumps have different line-of-sight velocities, which partly contributes to the remarkably large velocity dispersion observed in our ACA study.
Furthermore, the CO gas excitation traced by CO(2-1)/CO(1-0) ratio is low ($\sim$0.36 in brightness temperature units), implying that the Bridge is a relatively low-density region \citep{Emonts_2025ApJ}.
This suggests that the formation of GMCs is inhibited in this region.

In the molecular gas ridge (Ridge-N/Ridge-S), previous studies reported that this region tends to suppress star formation as described in Section~\ref{sec:compareSFE}, and several physical scenarios have been proposed to explain how shocks may inhibit (prevent) cloud formation and star formation within clouds.
\citet{Guillard_2009A&A} proposed that the clouds experiencing shock waves can interact with background low-density gas at different relative velocities, leading to cloud fragmentation. 
Such interactions may produce molecular fragments that are neither sufficiently massive nor long-lived to undergo gravitational collapse, thereby suppressing star formation. 
Another possibility is suggested by \citet{Xu+25}; the reason for the suppression of star formation in the Ridge is that if the pre-shock IGM is a low density and diffuse gas such as a warm neutral or ionized medium, the shock may produce strong vorticity, which inhibits cloud contraction and suppresses star formation.
In this context, \citet{Kobayashi_2022ApJ} performed high-resolution numerical simulations of shocked warm neutral medium. They found that velocity dispersion introduced in clouds becomes dominated by solenoidal-mode, which further suppresses high-density structure in clouds to form stars.
The interpretation by \citet{Xu+25} would be plausible if the velocity dispersion in the Ridge reflects a situation in which the velocity dispersions of individual molecular clouds are large, while the line-of-sight velocity differences among the clouds are small.
The causes of the suppression of star formation in those regions and their relation to the velocity dispersion of GMCs will be investigated with higher-angular resolution observations.

The region labeled NGC~7318B shows a large velocity dispersion (65~$\mathrm{km\,s^{-1}}$) compared to other moderate SFE regions ($\lesssim$39~$\mathrm{km\,s^{-1}}$).
The CO spectrum of NGC~7318B shows a redshifted wing (see Figure~\ref{fig:SQmap+profile}).
This feature is consistent with the ACA CO(2--1) observations reported by \citet{Emonts_2025ApJ}, who suggested that this emission originates from either pre-existing gas or molecular gas that has been decelerated relative to the intruder galaxy NGC~7318B. 
The presence of such kinematically distinct components within a small spatial scale likely leads to the large velocity dispersion observed in this region.

\subsection{Origin of molecular gas in tidal tails}
\label{sec: tails}

As mentioned above, we found several molecular gas components in the tidal tails, far from the main galaxy disks in SQ.
How such molecular gas, isolated in intergalactic space, is formed remains an open question.
In this subsection, we discuss the origin of SQ-B and Clump~1--4 by comparing the distributions of CO and H\,{\sc i}.

SQ-B is a prominent intergalactic star-forming region and has therefore been the target of some CO observations \citep[e.g.,][]{braine_abundant_2001,Lisenfeld_2002AA,Lisenfeld_2004AA,Maeda_2025ApJ}. 
These studies have shown that CO emission is spatially and kinematically coincident with H\,{\sc i} and extends beyond the sites of recent star formation, leading to the widely accepted interpretation that the molecular gas in SQ-B formed in situ from H\,{\sc i} stripped from NGC~7319 \citep{braine_abundant_2001,Lisenfeld_2002AA}. 
Our wide-area CO map shown in Figure 2 also supports this scenario.
However, we additionally detect a faint CO component without any associated H\,{\sc i} emission at the western edge of SQ-B, close to NGC~7319, as shown in the top right panel of Figure~\ref{fig:SQmap+profile}. 
This feature, which was also noted by \citet{Maeda_2025ApJ}, suggests that part of the molecular gas in SQ-B may have been directly stripped from NGC~7319, likely together with the stars that form the optical inner tail \citep{Fedotov2011AJ}.

Three CO clumps, Clump~1--3, are spatially and kinematically coincident with the H\,{\sc i} gas \citep{Williams_2002AJ}, suggesting that they likely formed in situ.
Our CO observation revealed that molecular-gas rich and molecular-gas poor regions coexist within the H\,{\sc i} tail.
In addition, the SFEs differ by orders of magnitude among Clump~1--3 while they have similar velocity dispersions (10 -- 15\,km s$^{-1}$).
Such an inhomogeneous molecular gas distribution and the variation in SFEs may reflect spatial variations in the local tidal field.
Numerical simulations demonstrate that compressive and extensive tidal modes can coexist within the tidal structures \citep{Renaud_2009ApJ}.
Compressive tides can enhance external pressure and promote the formation of dense molecular gas, whereas extensive tides tend to keep the gas diffuse \citep{Renaud_2014MNRAS}.
Within this framework, Clump~1--3 may correspond to locations where compressive tides locally act on the H\,{\sc i} gas, leading to the increase in gas density and subsequent conversion into molecular gas.
As shown by \citet{Renaud_2015MNRAS}, the strength of compressive tides in tidal tails depends on the combined gravitational potentials of the dark matter halos, the galaxies, and the tails themselves, as well as on the morphology of the tails.
Its variations are expected to influence the SFEs within the tails.
Weaker compressive tides in Clump~2 than Clump~1 and Clump~3 inhibit the enhancement of the dense gas fraction, resulting in low SFE even when the velocity dispersion is small.

The origin of Clump~4 is unclear.
Although this clump is partially overlapping with a foreground galaxy NGC~7320 along the line of sight, no counterparts are detected at other wavelengths, including H$\alpha$ \citep{Duarte_Puertas_2019AA,Duarte_Puertas_2021AA}, H\,{\sc i} \citep{Williams_2002AJ}, optical, or JWST infrared emission.
Clump~4 is also offset from both the inner and outer tidal tails (Section~\ref{sec: clumps}). 
Given the lack of associated H\,{\sc i} and any evidence for past or ongoing star formation, it is unlikely that Clump~4 has converted almost all of the H\,{\sc i} into molecular gas in situ due to local pressure or other processes.
We therefore interpret Clump~4 as molecular gas that was directly stripped from NGC~7319 because the radial velocity of Clump~4 ($\sim$6600~$\mathrm{km\,s^{-1}}$) is similar to the systemic velocity of NGC~7319 \citep[6740~$\mathrm{km\,s^{-1}}$;][]{Aoki_1996AJ}.
\citet{Bournaud_2006A&A} showed in numerical simulations that the survival of tidal dwarf–like objects depends strongly on where they form along the tidal tail; only objects forming near the tail tip can remain bound and long-lived ($\ge$2~Gyr), whereas objects forming at smaller radii are expected to fall back to the parent galaxy or be disrupted within a few 10$^8$ years.
Clump~4 lies closer to the member galaxies than the other clumps, and if it formed through tidal stripping during the formation of the inner tail, whose age is estimated to be $\sim$150--200~Myr \citep{Fedotov2011AJ}, this timescale is consistent with the continued presence of this clump stripped from the parent galaxy. 

Considering no sign of either past or current star formation and significantly larger velocity dispersion than other CO structures in the tail regions, the dynamical state of the molecular gas in Clump~4 is also explained by the turbulent motions and/or the difference in the line-of-sight velocity between clouds as discussed in Section~\ref{sec:SFE_vs_sigma}.
Further high-angular resolution observations are required to investigate the small-scale kinematics and to reveal the factors that control the difference in star formation activity among CO clumps and the dynamical processes driving star formation in the tidal tails.

\section{Conclusions}
\label{sec: conclusion}
 We conducted ALMA ACA (7\,m+TP) CO(1–0) observations covering the entire SQ system, providing the first large-scale, spatially resolved molecular gas map of a compact galaxy group. Our wide-area CO mapping ($137\,\mathrm{kpc}\times119\,\mathrm{kpc}$) at a spatial resolution of $5.85\times5.48\,\mathrm{kpc}$ resolves the major molecular gas structures, including the tidal tail.
 Our observation reveals some newly discovered molecular gas features in the tidal tail and significant variations in star formation activity and turbulence across the system.
Our main results are as follows:
\begin{enumerate}

   \item The molecular gas in SQ spans a wide velocity range ($\sim$1400~$\mathrm{km\,s^{-1}}$) and can be separated into three kinematically distinct components: a low-velocity component associated with NGC~7318B, a mid-velocity component connecting NGC~7318B and NGC~7319 through diffuse molecular gas outside the galaxies as a ridge component, and a high-velocity component that is widely distributed across NGC~7319, the Bridge, the tidal tail, and SQ-A (Figure~\ref{fig:SQmap+profile}).
   
   \item We identified an elongated molecular gas structure along the inner tidal tail, including SQ-B, together with four discrete CO clumps (Clump~1-4). Clump~1–3, located in the northern part of the inner tail, have masses of order $10^7\,M_\odot$ and small velocity dispersions ($<20\,\mathrm{km\,s^{-1}}$), whereas Clump~4, located between the inner and outer tails, is more massive ($>10^8\,M_\odot$) and exhibits a larger velocity dispersion ($\sim33~\mathrm{km\,s^{-1}}$) (Section~\ref{sec: clumps}).
    
    \item Using SFRs derived from H$\alpha$ emission of H\,{\sc ii} regions, we found that SFEs in SQ span $\sim2.2$\,dex (Figure~\ref{fig:KSlaw+SFE_sigma}).
    While several regions (the eastern arm of NGC~7318B, Northern Extension, SQ-A, SQ tip, SQ-B, Clump~1, and Clump~3) show SFEs comparable to those of nearby galaxies (0.6--4.0\,Gyr$^{-1}$), the molecular gas ridge, Bridge, Clump~2, and the northern area of NGC~7319 exhibit strongly suppressed SFEs (0.023--0.072\,Gyr$^{-1}$), at least an order of magnitude lower (Section~\ref{sec:compareSFE}).
    
    \item We found a clear negative correlation between the SFE and CO velocity dispersion (Figure~\ref{fig:KSlaw+SFE_sigma}). The large velocity dispersions observed in low SFE regions may reflect strong large-scale turbulence suppressing the formation of GMCs, large internal motions preventing the gravitational collapse of GMCs, and/or high-speed cloud-cloud collisions (Section~\ref{sec:SFE_vs_sigma}).

    \item Comparison with the H\,{\sc i} distribution shows that while most of the molecular gas located in SQ-B spatially overlaps with H\,{\sc i}, the CO components extending along the inner tidal tail partly lack an H\,{\sc i} counterpart, suggesting that not all of the molecular gas originates from in situ formation. 
    Clump~1–3 are spatially and kinematically associated with H\,{\sc i} and are therefore likely formed in situ, whereas Clump~4 shows no association with H\,{\sc i} and no evidence of past or ongoing star formation, making direct stripping from NGC~7319 a more plausible origin (Section~\ref{sec: tails}).

\end{enumerate}    
    
In conclusion, our results suggest that turbulence plays a significant role in regulating star formation in interacting systems.
However, high-angular resolution observations are required for investigating the physical origin connecting star formation with turbulence on the GMC scale or smaller.
It is also important to examine the difference in GMC-scale turbulence between interacting systems and isolated disk galaxies.
In addition, new insights may be gained by conducting different molecular line observations to understand the physical state of the gas, such as molecular gas density.

\begin{acknowledgments}
We would like to thank the anonymous referee for the useful comments, which helped improve the manuscript.
This paper makes use of the following ALMA data:  ADS/JAO.ALMA $\#$2023.1.01101.S. ALMA is a partnership of ESO (representing its member states), NSF (USA), and NINS (Japan), together with NRC (Canada), MOST and ASIAA (Taiwan), and KASI (Republic of Korea), in cooperation with the Republic of Chile. The Joint ALMA Observatory is operated by ESO, AUI/NRAO, and NAOJ. 
This work was supported by JSPS KAKENHI (grant Nos. JP23K13142, JP23H00129, JP23K20035, JP24H00004, JP24KJ1904, JP25K07371, and JP25K23396).
This work was supported in part by a University Research Support Grant from the National Astronomical Observatory of Japan (NAOJ).
\end{acknowledgments}

\appendix

\restartappendixnumbering
\section{Faint CO emission in tail regions}
We discovered four discrete CO clumps (Clumps~1--4) in the tail region. 
However, these structures are detected under relatively strict masking criteria, and additional faint molecular gas components are possibly present in the tails.
In this appendix, we present the integrated intensity map constructed without applying any mask in order to search for such faint emission. 

\begin{figure}[t!]
 \begin{center}
   \includegraphics[width=95mm]{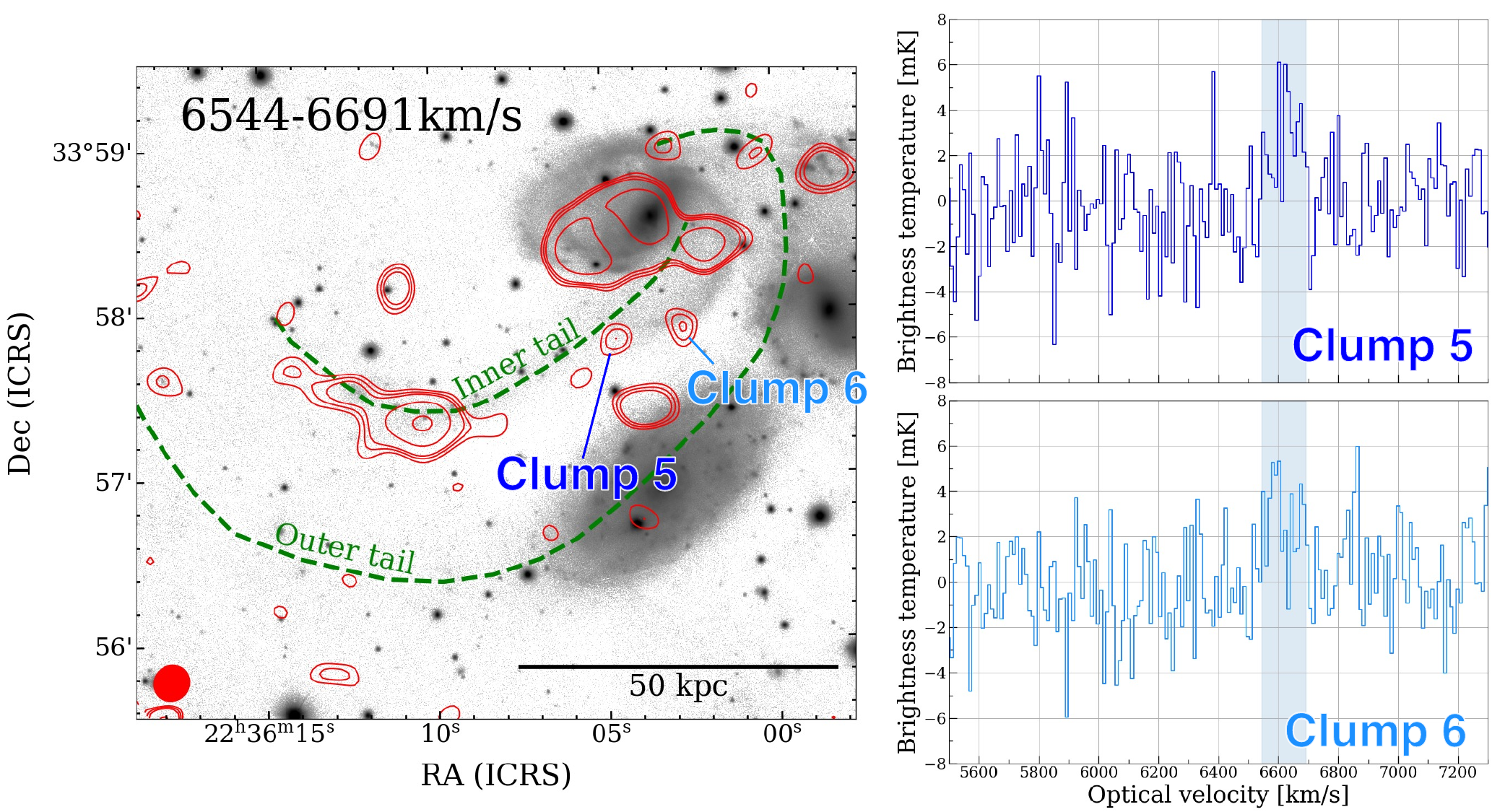} 
 \end{center} 
\caption{
(Left) Integrated intensity map of the non-masked CO(1--0) cube with a velocity resolution of 10~$\mathrm{km\,s^{-1}}$, constructed over the optical velocity range of 6544--6691~$\mathrm{km\,s^{-1}}$, where significant CO emission is detected in the tail region. 
The map is overlaid on the Pan-STARRS $R$-band image. 
The green dashed lines indicate the inner and outer tails. 
Contours are drawn at 3, 4, 5, 10, and 20$\sigma_{int}$, where $\sigma_{int} = 0.1$~K~$\mathrm{km\,s^{-1}}$ is the noise level of the integrated intensity.
(Right) Region-averaged spectra extracted within the regions enclosed by the lowest contour level for Clump~5 (blue) and Clump~6 (light blue), respectively, as shown in the left panel.
The blue bands indicate the velocity range used to construct the integrated intensity map shown in the left panel.
}\label{fig:appendixA}
\end{figure}

The left panel of Figure~\ref{fig:appendixA} shows the integrated intensity map derived from the non-masked CO(1--0) cube with a velocity resolution of 10~$\mathrm{km\,s^{-1}}$.
The integration is performed over the velocity range of 6544--6691~$\mathrm{km\,s^{-1}}$, where the significant emission in the tail is detected.
Two tentative CO clumps are newly identified only in the non-masked map (hereafter Clumps~5 and 6). 
The right panels show the region-averaged spectra within the area enclosed by the lowest contour level for Clumps~5 and 6. 
While the emission is not significant due to sensitivity limitations, the spectra suggest the possible presence of weak CO components within the velocity range used for the integration (indicated by the blue bands).
These results suggest that additional low-surface-brightness CO emission is likely present in the tail region but remains undetected due to sensitivity limitations.

\clearpage





%
\facilities{ALMA}

\software{CASA \citep{CASA_2022PASP}, Astropy \citep{astropy_2018}, APLpy \citep{Robitaille_2012}}



\bibliography{SQ}
\bibliographystyle{aasjournalv7}



\end{document}